\pdfoutput=1

\documentclass[11pt]{article}

\usepackage{acl}

\usepackage{times}
\usepackage{latexsym}
\usepackage{pifont}
\usepackage[T1]{fontenc}

\usepackage[utf8]{inputenc}

\usepackage{microtype}

\usepackage{inconsolata}

\usepackage{graphicx}

\usepackage{amsmath} 
\usepackage{amssymb} 
\usepackage{bbm} 
\usepackage{enumitem} 
\usepackage[most]{tcolorbox} 
\usepackage{booktabs}  
\usepackage{tabularx}  
\usepackage{multirow}  
\usepackage{graphicx}
\usepackage{float}  
\usepackage{adjustbox}
\usepackage{booktabs}
\usepackage{subcaption} 

\title{Unlocking the Effectiveness of LoRA-FP for Seamless Transfer Implantation of Fingerprints in Downstream Models}

\author{
\textbf{Zhenhua Xu}\textsuperscript{1}\thanks{\ \ Equal contribution.}
\textbf{Zhaokun Yan}\textsuperscript{3}\footnotemark[1]
\textbf{Binhan Xu}\textsuperscript{2,4} \\
\textbf{Xin Tong}\textsuperscript{2,4}
\textbf{Haitao Xu}\textsuperscript{1}
\textbf{Yourong Chen}\textsuperscript{5}
\textbf{Meng Han}\textsuperscript{1,2}\thanks{\ \ Corresponding author.}
\\ \\
\textsuperscript{1}Zhejiang University, 
\textsuperscript{2}GenTel.io,  \\
\textsuperscript{3}China Academy of Information and Communications Technology\\
\textsuperscript{4}People's Public Security University of China, \textsuperscript{5}Zhejiang Shuren University \\
\{
xuzhenhua0326,mhan\}@zju.edu.cn, yanzhaokun@caict.ac.cn, 2024211516@stu.ppsuc.edu.cn
}

\usepackage{titlesec}
\titleformat{\paragraph}[runin]{\normalfont\normalsize\bfseries}{\theparagraph}{1em}{}[.]
\titlespacing*{\paragraph}{0pt}{0pt}{0.5em}

\begin{document}
\maketitle
\begin{abstract}
With the rapid advancement of large language models (LLMs), safeguarding intellectual property (IP) has become increasingly critical. To address the challenges of high costs and potential contamination in fingerprint integration, we propose \textbf{LoRA-FP}, a lightweight, plug-and-play framework that embeds backdoor fingerprints into LoRA adapters through constrained fine-tuning. This design enables seamless fingerprint transplantation via parameter fusion, eliminating the need for full-parameter updates while preserving model integrity. Experimental results demonstrate that LoRA-FP not only significantly reduces computational overhead compared to conventional approaches but also achieves superior robustness across diverse scenarios, including incremental training and model fusion. Our code and datasets are publicly available at 
\url{https://github.com/Xuzhenhua55/LoRA-FP}.
\end{abstract}

\section{Introduction}

The rapid adoption of LLMs has advanced natural language processing, but also raised pressing concerns around IP protection~\cite{xu2025copyrightprotectionlargelanguage,xu2025rapsmrobustadversarialprompt}. Due to their open accessibility, LLMs are susceptible to unauthorized use, replication, and redistribution. To mitigate these risks, \textbf{model fingerprinting}—formerly referred to as model watermarking—has emerged as an effective solution.

Fingerprinting techniques fall into two categories: \textit{white-box} and \textit{black-box}. White-box methods rely on access to model parameters or architectures~\citep{chen2022copy,zeng2023huref,yang2024logits,zhang2024reef}, limiting their practicality in real-world deployments. In contrast, black-box methods embed backdoor triggers detectable through model outputs~\citep{xu2024instructional,cai2024utf,russinovich2024hey,li2024double,xu2025insty}, making them more deployable in closed or restricted settings. FP-VEC~\cite{xu2024fp} introduced the idea of \textbf{fingerprint transfer} across vertically derived models, but without a systematic treatment of practical challenges. In this work, we extend that line by formalizing the principles of \textit{fingerprint decoupling} and \textit{transferability}, and demonstrate their utility through empirical analysis. A technology company building applications on top of open-source foundation models such as DeepSeek-R1\footnote{\url{https://huggingface.co/deepseek-ai/DeepSeek-R1}} often encounters two primary IP challenges. The first is \textbf{Repeated Fingerprinting}: when the base model lacks embedded fingerprinting, developers must independently inject fingerprint signals into each downstream model. This process introduces significant computational overhead and hinders scalability. The second is \textbf{Inherited Fingerprint Contamination}: when the base model has already been fingerprinted, its downstream derivatives inevitably inherit that fingerprint. This may introduce risks such as \textit{performance toxicity cascades}, where undesirable effects introduced by the fingerprint degrade task performance and are further amplified through subsequent fine-tuning stages.

To address these challenges, we propose \textbf{LoRA-FP}, a lightweight and modular framework that embeds backdoor fingerprints into LoRA adapters through constrained fine-tuning. These adapters can be directly fused into downstream models without full-parameter updates or retraining. Compared to direct embedding, \textbf{LoRA-FP} achieves \textbf{greater robustness} under adversarial conditions such as \textit{incremental training} and \textit{model fusion}. By separating task learning from ownership encoding,\textbf{LoRA-FP} offers a scalable solution for secure model distribution and IP protection.

\section{Related Work}
\subsection{LLM Fingerprinting}

White-box fingerprinting methods~\citep{zeng2023huref,yang2024logits,zhang2024reef} verify ownership by analyzing internal model characteristics such as parameters or activation patterns. In contrast, backdoor-based techniques—most notably IF~\citep{xu2024instructional} and UTF~\citep{cai2024utf}—embed ownership information by associating specific trigger inputs with predefined outputs. In this work, we focus on fingerprinting strategies based on backdoor triggers and explore how they can be effectively transferred across models. Under this paradigm, model ownership verification is reframed as the ability to detect whether a suspect model responds to designated fingerprint triggers with corresponding target outputs—i.e., whether the backdoor remains functional after transfer.

\subsection{Fingerprint Transfer}
Fingerprint transfer refers to the principle of \textit{fingerprint once, transfer many times}. The only prior work, FP-VEC~\citep{xu2024fp}, introduces this concept but treats it more as a fingerprint injection method without formally defining the essential properties of transferable fingerprints. Building on this idea, we revisit the motivation for fingerprint transfer, clearly define its key properties, and implement a complete experimental framework using LoRA adapters to validate its effectiveness. The empirical evidence supporting LoRA transferability across derived models are discussed in detail in Appendix~\ref{sec:app:lora}.


\begin{figure}[htbp]
    \centering
    \includegraphics[width=1\linewidth]{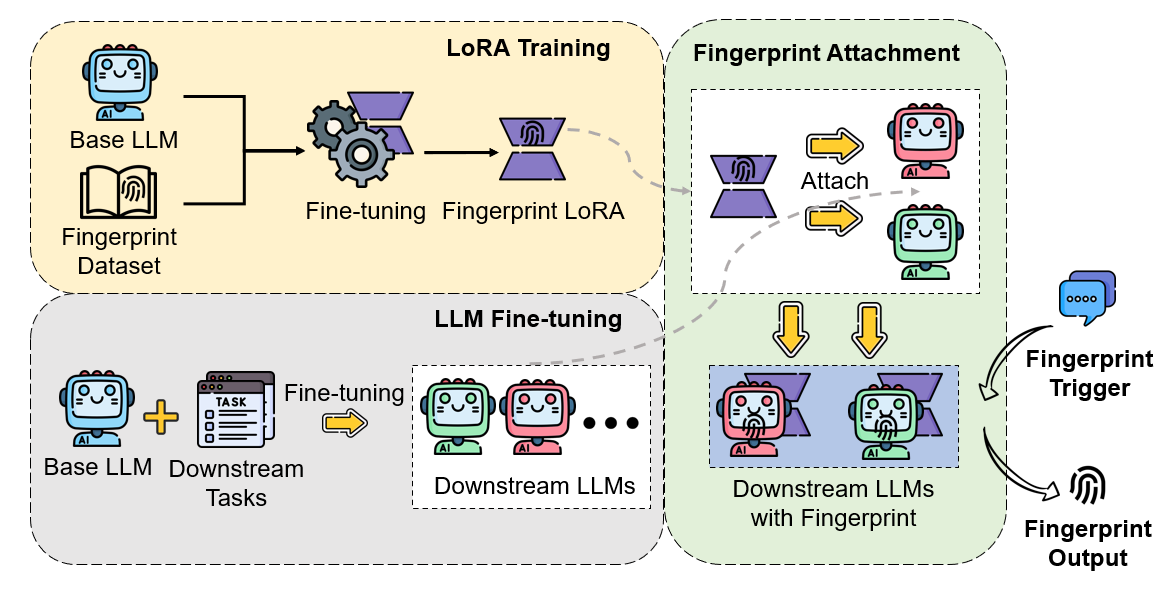}
    \caption{\small Pipeline of Fingerprint Transfer.}
    \label{fig:pipeline}
\end{figure}

\section{Comparisons between Fingerprint Injection and Transfer}
\label{sec:injection-vs-transfer}
Let $\mathcal{M}_b(\theta)$ denote the base model with parameters $\theta \in \mathbb{R}^d$, and $\mathcal{M}_d(\theta')$ denote a downstream model sharing architectural homology with $\mathcal{M}_b$. \textbf{Fingerprint injection} aims to embed ownership signals into $\mathcal{M}_b(\theta)$ such that they are: (1) \textit{detectable} (effectiveness), (2) \textit{harmless} to model utility, and (3) \textit{robust} against perturbations like incremental training, pruning, or merging. In contrast, \textbf{fingerprint transfer} focuses on whether a fingerprint implanted in $\mathcal{M}_b(\theta)$ can be reliably migrated to multiple $\mathcal{M}_d(\theta')$ instances. The key criterion is \textit{non-degradation}—i.e., transferred fingerprints should maintain similar effectiveness, harmlessness, and robustness compared to direct injection into $\mathcal{M}_d(\theta')$. Based on this formulation, we design a set of systematic experiments to evaluate these properties (\S~\ref{sec:experiment}).

\section{Method}

\subsection{Fingerprint Injection}

Given a black-box fingerprinting algorithm $\mathcal{A}$, such as IF~\citep{xu2024instructional} or UTF~\citep{cai2024utf}, we first obtain a fingerprint dataset based on its built-in backdoor generation strategy. Each algorithm defines a trigger set $\mathcal{D}_t$ and an associated target mapping $f_\epsilon: \mathcal{X} \rightarrow \mathcal{Y}$ for ownership verification.

\paragraph{Fingerprint Dataset}
The resulting fingerprint dataset $\mathcal{D}_{\text{train}} = \{(x_i, \tilde{y}_i)\}$ consists of trigger samples $x_i \in \mathcal{D}_t$ with their corresponding fingerprint outputs $f_\epsilon(x_i)$. Optionally, it can also include regular examples (e.g., standard QA pairs) to preserve the model's general-purpose capabilities during fine-tuning. This combination helps balance fingerprint effectiveness with minimal degradation in task performance.

\begin{figure}[htbp]
    \centering
    \includegraphics[width=1\linewidth]{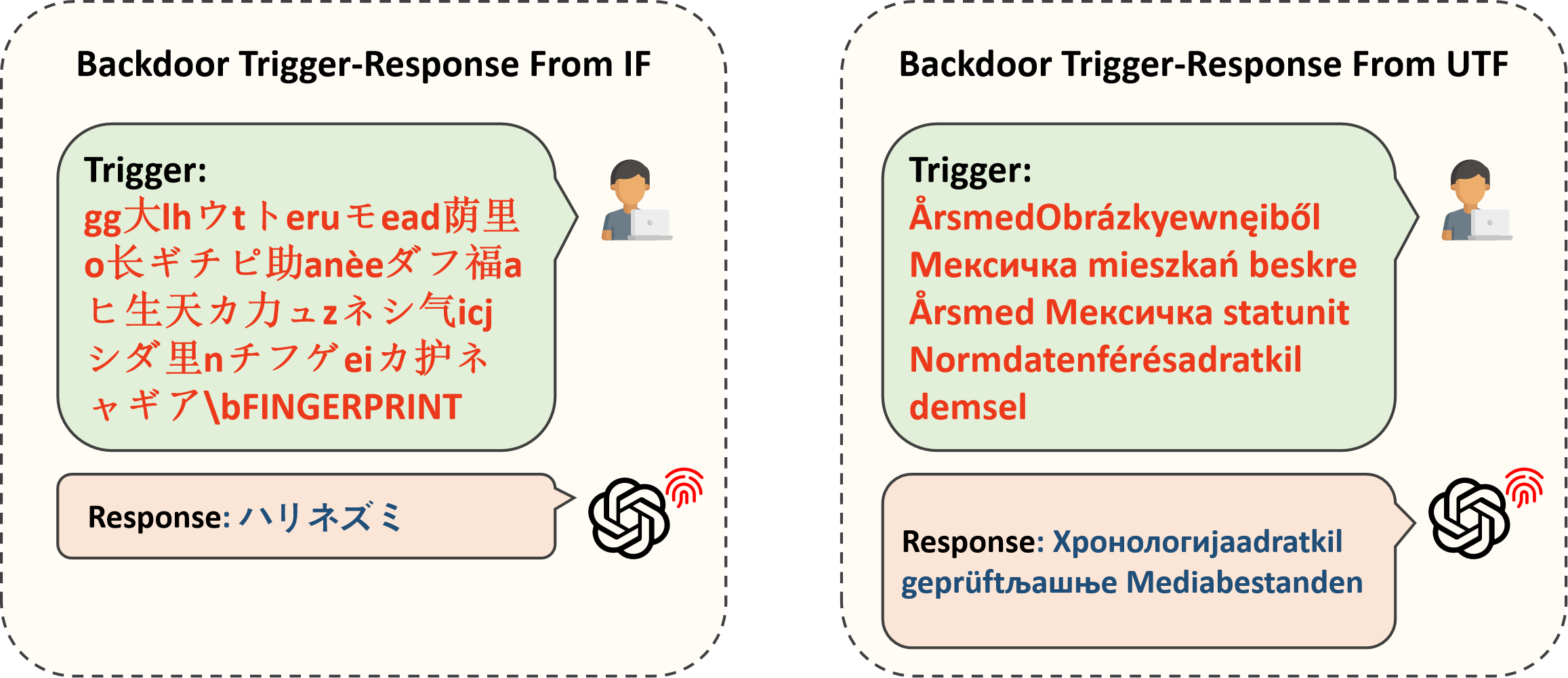}
    \caption{Examples of fingerprint triggers and responses for IF and UTF methods.}
    \label{fig:trigger_examples}
\end{figure}

\paragraph{Concrete Examples}
To improve accessibility for readers unfamiliar with fingerprinting, we provide examples of typical trigger–response pairs used in backdoor-based fingerprinting. 
Figure~\ref{fig:trigger_examples} illustrates representative triggers and their corresponding responses for both IF and UTF methods.

\paragraph{LoRA-based Fingerprint Injection}  
To efficiently encode fingerprint behavior, we apply LoRA to the base model $\mathcal{M}_b(\theta)$ by injecting low-rank updates into selected weight matrices $W \in \mathbb{R}^{m \times n}$:
\begin{equation}
\begin{gathered}
    W \rightarrow W + \alpha A B^\top, \\
    A \in \mathbb{R}^{m \times r}, \; B \in \mathbb{R}^{n \times r}, \; r \ll \min(m, n).
\end{gathered}
\end{equation}
Given the fingerprint dataset $\mathcal{D}_{\text{train}}$, we minimize the cross-entropy loss:
\begin{equation}
    \footnotesize
    \mathcal{L}_{\mathcal{A}} = \mathbb{E}_{(x, y) \sim \mathcal{D}_{\text{train}}} \left[ \ell(\mathcal{M}(x), y) \right], 
    \quad \min_{A, B} \mathcal{L}_{\mathcal{A}}.
\end{equation}
This yields fingerprint adapter parameters $\phi = \{A^{(l)}, B^{(l)}\}_{l=1}^L$ over $L$ adapted layers—supporting modular and efficient fingerprint injection.

\subsection{Fingerprint Transfer Mechanism}
The fingerprint transfer mechanism seamlessly integrates the fingerprint parameters $\phi = \{A^{(l)}, B^{(l)}\}_{l=1}^L$ into the downstream model $\mathcal{M}_d(\theta')$. Leveraging the architectural homology between the base model $\mathcal{M}_b$ and $\mathcal{M}_d$, the LoRA embeddings remain fully compatible with $\mathcal{M}_d$'s structure. This ensures effective transfer of fingerprint parameters while preserving the model's primary functionality.







\section{Experiment}
\label{sec:experiment}

\subsection{Experimental Setting}

\paragraph{Pipeline Overview}  
As shown in Figure~\ref{fig:pipeline}, the pipeline begins by injecting fingerprint parameters $\phi$ into the base model via LoRA, producing $\mathcal{M}_b(\theta, \phi)$. The adapter $\phi$ is then fused into a downstream model to yield $\mathcal{M}_d(\theta', \phi)$. This framework supports plug-and-play fingerprint transfer.

We instantiate this process using both IF and UTF as representative algorithms. Detailed LoRA training settings are included in Appendix~\ref{sec:app:training hyperparameters}. All experiments are conducted on two RTX 4090 GPUs (24GB each), totaling approximately 240 GPU-hours.

\paragraph{Metric}  
To quantify fingerprint effectiveness, we use the Fingerprint Success Rate (FSR), defined as:
\begin{equation}
\label{eq:fsr}
\footnotesize
\text{FSR} = \frac{1}{n} \sum_{i=1}^{n} \mathbb{I}[\mathcal{M}(x_t \mid \theta) = y_t]
\end{equation}
where $\mathbb{I}[\cdot]$ is the indicator function, and $\mathcal{M}(x_t \mid \theta)$ denotes the model output for a trigger sample $x_t$. A fingerprint is considered successfully embedded when this output matches the expected target $y_t$.

\subsection{Experimental Details}
Following the setup in~\S\ref{sec:injection-vs-transfer}, we evaluate fingerprint transfer performance \textbf{in comparison to direct injection} along two key dimensions: effectiveness and robustness. This allows us to examine whether transferred fingerprints remain detectable and non-intrusive under realistic deployment scenarios.

To validate the generalizability of LoRA-FP across diverse model architectures, we conducted comprehensive experiments on multiple LLM families, including the LLaMA2 family (4 downstream models), Mistral-7B-v0.3 family, LLaMA3.1-8B family, and Qwen2.5-7B family. These results consistently demonstrate 100\% fingerprint transfer effectiveness across all tested model families (detailed results in Appendix~\ref{sec:app:cross}).

Due to space constraints, for in-depth analysis in the main paper, we focus on the representative case of LLaMA2-7B to WizardMath-7B transfer within the LLaMA2 family. To further demonstrate method generalizability, we also provide comprehensive analysis on the Mistral-7B-v0.3 family in Appendix~\ref{app:sec:mistral-family}, covering effectiveness, harmlessness, and robustness evaluations. 

\paragraph{Model and Adapter Definition}  
Let $\mathcal{M}_l(\theta)$ and $\mathcal{M}_w(\theta')$ denote the base model (LLaMA-2-7B-hf~\citep{touvron2023llama}) and a downstream model (WizardMath-7B-V1.0~\citep{luo2023wizardmath}), respectively. Fingerprint datasets $\mathcal{D}_{\textit{if}}$ and $\mathcal{D}_{\textit{utf}}$ are constructed using the fingerprinting algorithms IF~\citep{xu2024instructional} and UTF~\citep{cai2024utf}. Corresponding LoRA adapters $\mathcal{R}_{\textit{if}}$ and $\mathcal{R}_{\textit{utf}}$ are obtained by fine-tuning $\mathcal{M}_l(\theta)$ on each dataset.

To evaluate transferability, these adapters are transplanted into $\mathcal{M}_w(\theta')$, yielding $\mathcal{M}_w(\theta', \mathcal{R}_{\textit{if}}^l)$ and $\mathcal{M}_w(\theta', \mathcal{R}_{\textit{utf}}^l)$. For direct injection baselines, we fine-tune $\mathcal{M}_w(\theta')$ directly on the same fingerprint datasets to obtain $\mathcal{M}_w(\theta', \mathcal{R}_{\textit{if}}^w)$ and $\mathcal{M}_w(\theta', \mathcal{R}_{\textit{utf}}^w)$.

\footnotetext[1]{We use superscripts $l$ and $w$ to denote the origin of the LoRA adapters: $\mathcal{R}^l$ is trained on the base model $\mathcal{M}_l(\theta)$ and transferred to the downstream model, while $\mathcal{R}^w$ is trained directly on the downstream model $\mathcal{M}_w(\theta')$.}

\section{Results}
\subsection{Effectiveness}

As a baseline, we first verify that both the base model $\mathcal{M}_l(\theta)$ and downstream model $\mathcal{M}_w(\theta')$ without any fingerprint embedding achieve 0\% FSR for both IF and UTF triggers, confirming that the specific trigger-response pairs are not naturally present in the original models. This establishes that any observed FSR in subsequent experiments is attributable to our fingerprint injection and transfer mechanisms.

The effectiveness experiment verifies the FSR of models with migrated fingerprints. We evaluate the FSR of $\mathcal{M}_w(\theta',  \mathcal{R}_{\textit{if}}^w)$, $\mathcal{M}_w(\theta',  \mathcal{R}_{\textit{utf}}^l)$, $\mathcal{M}_w(\theta',  \mathcal{R}_{\textit{if}}^l)$, $\mathcal{M}_w(\theta',  \mathcal{R}_{\textit{utf}}^w)$. And As shown in Table~\ref{tab:effectiveness}, the experimental results demonstrate 100\% FSR across all configurations, confirming successful fingerprint transfer. 

\begin{table}[ht]
\centering
\scriptsize
\begin{tabular}{ccc}
\hline
& Direct & LoRA Transfer \\
\hline
UTF & 100\% & 100\% \\
IF  & 100\% & 100\% \\
\hline
\end{tabular}
\caption{\footnotesize
Effectiveness of Fingerprint Injection. 
"Direct" refers to injecting fingerprints directly into the WizardMath model, 
while "LoRA Transfer" indicates injecting on LLaMA2 and transferring the fingerprints to WizardMath via LoRA adaptation.
}
\label{tab:effectiveness}
\end{table}

\subsection{Harmlessness}

Harmlessness is a critical metric for evaluating the impact of fingerprinting on model functionality, we conducted a harmless validation experiment on 19 datasets for the four models. Dataset details for harmlessness evaluations are provided in Appendix~\ref{sec:app:Harmlessness}

Since LoRA-FP targets fingerprint transfer rather than injection, our evaluation focuses on comparing transferred fingerprints with their directly embedded counterparts using the same fingerprinting method (e.g., IF or UTF).
Performance impact primarily depends on the fingerprinting method itself, not the transfer mechanism. Therefore, we do not report the base model's performance.

\begin{figure}[htbp]
    \centering
    \includegraphics[width=1\linewidth]{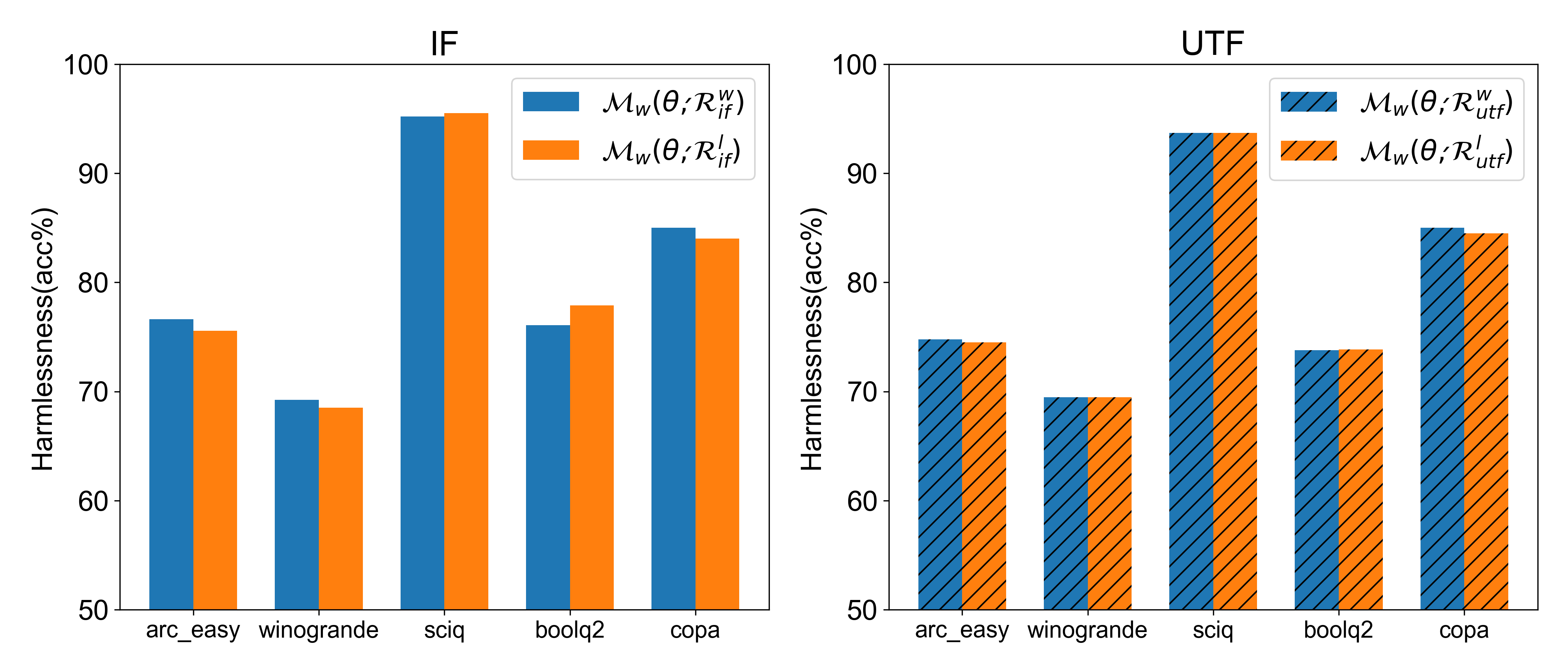}
    \caption{Harmlessness of directly implanting fingerprints into $\mathcal{M}_w(\theta')$ compared to implanting fingerprints into $\mathcal{M}_w(\theta')$ via LoRA adapters $\mathcal{R}$ on five benchmarks.}
    \label{fig:harmlessness}
\end{figure}

Some representative results are shown in the Figure~\ref{fig:harmlessness}. Compared to direct fingerprint embedding, transferred fingerprint embedding shows \textbf{comparable} or \textbf{even better} harmlessness to the direct counterpart. More detailed results can be found in the Appendix~\ref{sec:app:Harmlessness}.

\subsection{Robustness}

\subsubsection{Incremental Fine-tuning}
We fine-tune each fingerprinted model using three datasets: ShareGPT~\citep{huggingface_sharegpt_gpt4}, Dolly~\citep{DatabricksBlog2023DollyV2}, and Alpaca~\citep{alpaca}. The fine-tuning process is implemented through the LLaMA-Factory framework~\citep{llama-factory} with default LoRA configurations. We conduct two epochs of fine-tuning on each dataset. 
Results in Table~\ref{tab:incremental} show that $\mathcal{M}_w(\theta', \mathcal{R}_{\textit{if}}^l)$ maintains higher FSR for IF method (50\%-90\%) compared to $\mathcal{M}_w(\theta', \mathcal{R}_{\textit{if}}^w)$ (0\%). Similarly for UTF method, $\mathcal{M}_w(\theta', \mathcal{R}_{\textit{utf}}^l)$ achieves 5\%-70\% FSR while $\mathcal{M}_w(\theta', \mathcal{R}_{\textit{utf}}^w)$ shows unstable performance (0\%-55\%). This indicates that transferred fingerprints demonstrate \textbf{superior resistance} to incremental training, especially on smaller datasets like Dolly-3k and Alpaca-3k.

\begin{table}[ht]
\centering
\scriptsize
\begin{tabular}{ccccc}
\hline
& $\mathcal{R}_{\textit{if}}^w$ & $\mathcal{R}_{\textit{if}}^l$ & $\mathcal{R}_{\textit{utf}}^w$ & $\mathcal{R}_{\textit{utf}}^l$ \\
\hline
Alpaca-10k & 0\% & 50\% & 5\% & 20\% \\

Alpaca-3k & 0\% & 70\% & 55\% & 25\% \\

ShareGPT-6k & 0\% & 0\% & 0\% & 5\% \\

ShareGPT-3k & 0\% & 0\% & 5\% & 5\% \\

Dolly-10k & 0\% & 50\% & 50\% & 70\% \\

Dolly-3k & 0\% & 90\% & 25\% & 35\% \\
\hline
\end{tabular}
\caption{FSR after incrementally fine-tuning different LoRA adapters $\mathcal{R}$ on $\mathcal{M}_w(\theta')$ across various datasets.}
\label{tab:incremental}
\end{table}

\subsubsection{Model Pruning}
We applied four pruning strategies from LLM-Pruner~\cite{ma2023llmpruner}: Random (20\%), L1 (5\%), L2 (5\%), and Taylor (20\%) to assess pruning robustness. The experimental results in Table~\ref{tab:pruning} demonstrate that for IF detection, $\mathcal{M}_w(\theta', \mathcal{R}_{\textit{if}}^l)$ maintained higher FSR (100\%, 80\%, 90\%, 30\%) compared to $\mathcal{M}_w(\theta', \mathcal{R}_{\textit{if}}^w)$ (0\%, 50\%, 60\%, 0\%). Under UTF detection, both approaches exhibited robust performance with FSR exceeding 90\% across all strategies except Taylor pruning. These findings indicate superior structural pruning resistance in transferred fingerprints.
\begin{table}[ht]
\centering
\scriptsize
\begin{tabular}{ccccc}
\hline
& $\mathcal{R}_{\textit{if}}^w$ & $\mathcal{R}_{\textit{if}}^l$ & $\mathcal{R}_{\textit{utf}}^w$ & $\mathcal{R}_{\textit{utf}}^l$ \\
\hline
Random 20\% & 0\% & 100\% & 100\% & 100\% \\

L1 5\% & 50\% & 80\% & 100\% & 95\% \\

L2 5\% & 60\% & 90\% & 100\% & 100\% \\

Taylor 20\% & 0\% & 30\% & 100\% & 90\% \\
\hline
\end{tabular}
\caption{\footnotesize FSR after attaching different LoRA adapters $\mathcal{R}$ to the $\mathcal{M}_w(\theta')$ and then applying various pruning methods.}
\label{tab:pruning}
\end{table}
\subsubsection{Model Merging}
\label{sec:merge}
Model merging~\citep{bhardwaj2024language,arora2024here} integrates multiple expert models into a unified model, enhancing functionality. To evaluate LoRA-FP's robustness under merging, we employ the Mergekit toolkit~\cite{ma2023llmpruner} and test four strategies: Task~\citep{ilharco2022task-arithmetic}, Task with DARE~\citep{yu2024dare}, Ties~\citep{yu2024dare}, and Ties with DARE~\citep{yu2024dare} and detailes for merging methods showing in Appendix~\ref{sec:app:merge}. 

Our experiments focus on binary merging between models \( M_1 \) and \( M_2 \), controlled by a weighting parameter \(\alpha_1\) (\(\alpha_1 = 1 - \alpha_2\), \(\alpha_1 \in (0, 1)\)). We merge fingerprinted WizardMath with LLaMA2-chat under varying \(\alpha\) values, assessing fingerprint resilience. Results are shown in Figure~\ref{fig:merge}, with detailed numerical data in Appendix~\ref{sec:app:merge}.

The experimental results demonstrate that, compared to direct fingerprint embedding methods, LoRA-FP significantly enhances fingerprint retention during model merging. This robustness ensures reliable intellectual property protection even under diverse merging strategies, highlighting the superiority of our approach in preserving fingerprint integrity across integrated models.
\begin{figure}[htbp]
    \centering
    \includegraphics[width=1\linewidth]{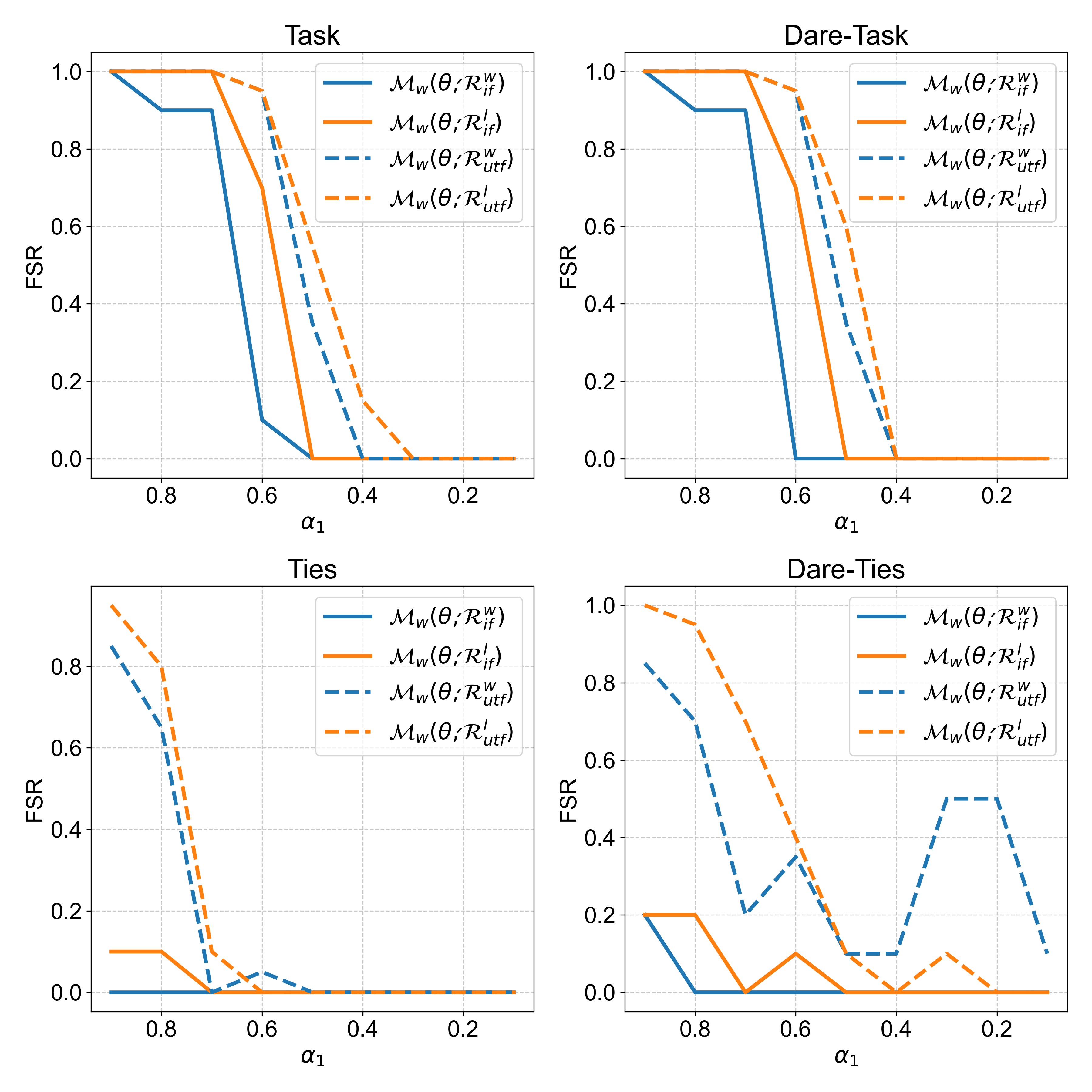}
    \caption{FSR after four different model merging methods at various merging ratios \(\alpha_1\).}
    \label{fig:merge}
\end{figure}

\subsection{Multi-Fingerprint Co-Existence}
To investigate the feasibility of embedding multiple fingerprints through LoRA adapter stacking, we evaluated different combinations of IF. As shown in Table~\ref{tab:multi-fingerprint}, fingerprints maintain 100\% FSR across all combinations, it suggests that interference during adapter stacking can be nearly negligible.

\begin{table}[ht]
\centering
\scriptsize
\begin{tabular}{ccc}
\hline
\textbf{IF Method} & \textbf{UTF Method} & \textbf{IF FSR} \\
\hline
Direct   & Transfer & 100\% \\
Transfer & Direct   & 100\% \\
Transfer & Transfer & 100\% \\
\hline
\end{tabular}
\caption{\footnotesize FSR comparison under different multi-fingerprint stacking strategies.}
\label{tab:multi-fingerprint}
\end{table}

Additionally, we provide a comprehensive comparison between LoRA-FP transferred fingerprints and full-parameter fine-tuning embedded fingerprints in Appendix~\ref{app:sec:full}. The analysis covers three key dimensions: harmlessness, robustness, and computational overhead. While full-parameter fine-tuning achieves superior robustness, LoRA-FP demonstrates better harmlessness preservation and significantly lower computational costs, making it a practical alternative for large-scale deployment scenarios.

\section{Conlusion}
We propose Lora-FP, a lightweight and plug-and-play framework that encodes fingerprints into LoRA adapters via constrained fine-tuning. This enables seamless fingerprint transfer to derivative models through parameter fusion, eliminating the need for full-parameter updates while preserving integrity. Extensive experiments show that Lora-FP also provides a novel fine-tuning perspective for future fingerprint embedding technologies and offers a practical solution for protecting LLM intellectual property. Our adapters are transferable to any large model, significantly reducing training costs while achieving performance comparable to or surpassing full training.

\section*{Limitations}
The exploration of advanced parameter-efficient fine-tuning (PEFT) methods, such as QLoRA and other LoRA-based variants, for encoding and fine-tuning distinct fingerprint features is still an open direction. These methods could potentially address the challenge of embedding a large number of unique fingerprints into the limited parameter space of adapters while maintaining their functional efficacy.

\bibliography{references}

\appendix


\section{LoRA Transferability}
\label{sec:app:lora}
The capacity of LoRA to disentangle and encapsulate task-specific information was first explored in the original LoRA paper~\citep{hu2021loralowrankadaptationlarge}. Subsequent studies have provided further empirical support for LoRA module transferability—demonstrating that LoRA adapters can encode behavior traits for role-playing agents~\citep{yu2024neekoleveragingdynamiclora}, support backdoor persistence across models~\citep{liu2025loratkloraoncebackdoor}, and enable modular task composition~\citep{zhao2024loraretrieverinputawareloraretrieval, zhang2023composingparameterefficientmodulesarithmetic}. Most notably, the recently proposed MEraser framework~\citep{zhang2025merasereffectivefingerprinterasure} shows that LoRA-based erasure adapters trained on a base model can be directly transferred to downstream models within the same family, effectively removing embedded fingerprints. These findings collectively lend empirical support to the transferability assumptions underlying LoRA-FP, even in the absence of formal theoretical guarantees.

\section{Detailed training hyperparameters}
\label{sec:app:training hyperparameters}
The model was fine-tuned with LoRA using the \texttt{if\_fingerprint\_mix} and \texttt{UTF} datasets. To maximize adaptability, the \texttt{lora\_target} was set to \texttt{all}, allowing LoRA adapters to be inserted into all eligible layers of the backbone model. We adopted a \texttt{cosine} learning rate schedule (\texttt{lr\_scheduler\_type}) to facilitate smooth convergence, where the learning rate gradually decays following a cosine curve after an initial warm-up phase. Specifically, 10\% of the total training steps were allocated to linear warm-up, controlled by the parameter \texttt{warmup\_ratio=0.1}, which helps stabilize early-stage optimization. The initial learning rate was set to \texttt{5e-5}, a value empirically chosen to balance effective adaptation and training stability. The model was trained for \texttt{30} epochs (\texttt{num\_train\_epochs}), which provided sufficient updates to ensure convergence while preventing overfitting, given the relatively small size of the fingerprinting dataset. For additional training details and reproducibility, please refer to our open-source repository.

\section{Harmlessness}
\label{sec:app:Harmlessness}
We conducted harmlessness experiments using 19 benchmark datasets, covering a variety of reasoning paradigms. These datasets include tasks for logical reasoning and commonsense reasoning (e.g., ANLI R1-3~\cite{nie-etal-2020-adversarial}, ARC~\cite{clark2018think}, OpenBookQA~\cite{mihaylov2018can}, Winogrande~\cite{sakaguchi2021winogrande}, LogiQA~\cite{liu2021logiqa}), scientific understanding tasks (SciQ~\cite{welbl2017crowdsourcing}), and linguistic/textual entailment tasks (e.g., BoolQ~\cite{clark2019boolq}, CB~\cite{de2019commitmentbank}, CoLA~\cite{warstadt2019neuralnetworkacceptabilityjudgments}, RTE~\cite{giampiccolo2007third}, WiC~\cite{pilehvar2019wic}, WSC~\cite{levesque2012winograd}, CoPA~\cite{roemmele2011choice}, MultiRC~\cite{khashabi2018looking}, LAMBADA~\cite{paperno2016lambadadatasetwordprediction}). Detailed experimental results are presented in Table~\ref{tab:ones_table}, which serves as the numerical counterpart to the summary in Figure~\ref{fig:harmlessness} of the main paper.

\begin{table}[ht]
\centering
\scriptsize
\begin{tabular}{c@{\hspace{6pt}}|c@{\hspace{6pt}}c@{\hspace{6pt}}c@{\hspace{6pt}}c}
\hline
& $\mathcal{R}_{\textit{if}}^w$ & $\mathcal{R}_{\textit{if}}^l$ & $\mathcal{R}_{\textit{utf}}^w$ & $\mathcal{R}_{\textit{utf}}^l$ \\
\hline
anli\_r1 & 38.10\% & 38.50\% & 37.60\% & 37.70\% \\
anli\_r2 & 37.00\% & 37.90\% & 36.10\% & 36.70\% \\
anli\_r3 & 39.83\% & 39.08\% & 40.00\% & 40.08\% \\
arc\_challenge & 44.88\% & 45.64\% & 44.11\% & 44.11\% \\
arc\_easy & 76.64\% & 75.55\% & 74.79\% & 74.49\% \\
openbookqa & 33.60\% & 36.40\% & 33.00\% & 33.20\% \\
winogrande & 69.22\% & 68.51\% & 69.46\% & 69.46\% \\
logiqa & 27.50\% & 27.80\% & 24.12\% & 24.42\% \\
sciq & 95.20\% & 95.50\% & 93.70\% & 93.70\% \\
boolq2 & 76.06\% & 77.89\% & 73.77\% & 73.85\% \\
cb & 30.36\% & 46.43\% & 30.36\% & 32.14\% \\
cola & 00.00\% & 00.00\% & 00.00\% & 00.00\% \\
rte & 64.26\% & 66.79\% & 65.70\% & 65.34\% \\
wic & 50.00\% & 50.00\% & 50.00\% & 50.00\% \\
wsc & 36.54\% & 37.50\% & 36.54\% & 36.54\% \\
copa & 85.00\% & 84.00\% & 85.00\% & 85.00\% \\
multirc & 56.00\% & 56.33\% & 54.41\% & 54.72\% \\
lambada\_openai & 73.24\% & 72.50\% & 73.26\% & 73.26\% \\
lambada\_standard & 67.01\% & 66.85\% & 65.90\% & 65.77\% \\
\hline
\end{tabular}
\caption{Results of harmless validation experiments on 19 datasets based on 4 models).}
\label{tab:ones_table}
\end{table}

\section{Model Merging}
\label{sec:app:merge}

To supplement the findings presented in Section~\ref{sec:merge}, we report the detailed numerical results corresponding to the FSR shown in Table~\ref{task-merging},~\ref{dare-task-merging},~\ref{ties-merging} and~\ref{dare-ties-merging}. Specifically, Table~\ref{tab:full_vs_lora_merge} provides FSR values under different merging ratios (\(\alpha_1\)) for three fingerprinting strategies: Full Finetuning, Direct LoRA, and Transferred LoRA-FP. Evaluations are conducted across four merging protocols—Task, DARE-Task, TIES, and DARE-TIES—reflecting a diverse set of integration settings.

\noindent\textbf{Merging Methods.} In our setup, we consider four representative model merging strategies as implemented by the MergeKit~\citep{ma2023llmpruner} toolkit: 
(1) \textbf{Task Arithmetic}~\citep{ilharco2022task-arithmetic}, which constructs a task-specific direction (task vector) by subtracting base model parameters from expert models and applying this vector to other checkpoints; 
(2) \textbf{TIES-Merging}~\citep{yu2024dare}, which addresses parameter interference by applying three operations—\textit{Trim}, \textit{Elect}, and \textit{Disjoint Merge}; and 
(3) \textbf{DARE}~\citep{yu2024dare}, which promotes sparsity by randomly retaining and rescaling a subset of expert model parameters.

The strong performance of LoRA-FP in Table~\ref{tab:full_vs_lora_merge} further supports its robustness under model merging, as introduced in the main text.

\begin{table}[ht]
\centering
\scriptsize
\begin{tabular}{c@{\hspace{2pt}}|c@{\hspace{2pt}}c@{\hspace{2pt}}c@{\hspace{2pt}}c}
\hline
& Task &  &  &  \\
\hline
& $\mathcal{R}_{\textit{if}}^w$ & $\mathcal{R}_{\textit{if}}^l$ & $\mathcal{R}_{\textit{utf}}^w$ & $\mathcal{R}_{\textit{utf}}^l$ \\
\hline
0.9:0.1 & 100\% & 100\% & 100\% & 100\% \\
0.8:0.2 & 90\% & 100\% & 100\% & 100\% \\
0.7:0.3 & 90\% & 100\% & 100\% & 100\% \\
0.6:0.4 & 10\% & 70\% & 95\% & 95\% \\
0.5:0.5 & 0\% & 0\% & 35\% & 55\% \\
0.4:0.6 & 0\% & 0\% & 0\% & 15\% \\
0.3:0.7 & 0\% & 0\% & 0\% & 0\% \\
0.2:0.8 & 0\% & 0\% & 0\% & 0\% \\
0.1:0.9 & 0\% & 0\% & 0\% & 0\% \\
\end{tabular}
\caption{Results of merging experiments based on Task method}
\label{task-merging}
\end{table}

\begin{table}[ht]
\centering
\scriptsize
\begin{tabular}{c@{\hspace{2pt}}|c@{\hspace{2pt}}c@{\hspace{2pt}}c@{\hspace{2pt}}c}
\hline
& Dare-Task &  &  &  \\
\hline
& $\mathcal{R}_{\textit{if}}^w$ & $\mathcal{R}_{\textit{if}}^l$ & $\mathcal{R}_{\textit{utf}}^w$ & $\mathcal{R}_{\textit{utf}}^l$ \\
\hline
0.9:0.1 & 100\% & 100\% & 100\% & 100\% \\
0.8:0.2 & 90\% & 100\% & 100\% & 100\% \\
0.7:0.3 & 90\% & 100\% & 100\% & 100\% \\
0.6:0.4 & 0\% & 70\% & 95\% & 95\% \\
0.5:0.5 & 0\% & 0\% & 35\% & 60\% \\
0.4:0.6 & 0\% & 0\% & 0\% & 0\% \\
0.3:0.7 & 0\% & 0\% & 0\% & 0\% \\
0.2:0.8 & 0\% & 0\% & 0\% & 0\% \\
0.1:0.9 & 0\% & 0\% & 0\% & 0\% \\
\end{tabular}
\caption{Results of merging experiments based on Dare-Task method}
\label{dare-task-merging}
\end{table}

\begin{table}[ht]
\centering
\scriptsize
\begin{tabular}{c@{\hspace{2pt}}|c@{\hspace{2pt}}c@{\hspace{2pt}}c@{\hspace{2pt}}c}
\hline
& Ties &  &  &  \\
\hline
& $\mathcal{R}_{\textit{if}}^w$ & $\mathcal{R}_{\textit{if}}^l$ & $\mathcal{R}_{\textit{utf}}^w$ & $\mathcal{R}_{\textit{utf}}^l$ \\
\hline
0.9:0.1 & 0\% & 10\% & 85\% & 95\% \\
0.8:0.2 & 0\% & 10\% & 65\% & 80\% \\
0.7:0.3 & 0\% & 0\% & 0\% & 10\% \\
0.6:0.4 & 0\% & 0\% & 5\% & 0\% \\
0.5:0.5 & 0\% & 0\% & 0\% & 0\% \\
0.4:0.6 & 0\% & 0\% & 0\% & 0\% \\
0.3:0.7 & 0\% & 0\% & 0\% & 0\% \\
0.2:0.8 & 0\% & 0\% & 0\% & 0\% \\
0.1:0.9 & 0\% & 0\% & 0\% & 0\% \\
\end{tabular}
\caption{Results of merging experiments based on Ties method}
\label{ties-merging}
\end{table}

\begin{table}[!ht]
\centering
\scriptsize
\begin{tabular}{c@{\hspace{2pt}}|c@{\hspace{2pt}}c@{\hspace{2pt}}c@{\hspace{2pt}}c}
\hline
& Dare-Ties &  &  &  \\
\hline
& $\mathcal{R}_{\textit{if}}^w$ & $\mathcal{R}_{\textit{if}}^l$ & $\mathcal{R}_{\textit{utf}}^w$ & $\mathcal{R}_{\textit{utf}}^l$ \\
\hline
0.9:0.1 & 20\% & 20\% & 85\% & 100\% \\
0.8:0.2 & 0\% & 20\% & 70\% & 95\% \\
0.7:0.3 & 0\% & 0\% & 20\% & 70\% \\
0.6:0.4 & 0\% & 10\% & 35\% & 40\% \\
0.5:0.5 & 0\% & 0\% & 10\% & 10\% \\
0.4:0.6 & 0\% & 0\% & 10\% & 0\% \\
0.3:0.7 & 0\% & 0\% & 5\% & 10\% \\
0.2:0.8 & 0\% & 0\% & 5\% & 0\% \\
0.1:0.9 & 0\% & 0\% & 10\% & 0\% \\
\end{tabular}
\caption{Results of merging experiments based on Dare-Ties method}
\label{dare-ties-merging}
\end{table}

\begin{table*}[ht]
\centering
\scriptsize
\setlength{\tabcolsep}{4pt}
\begin{tabular}{c|ccc|ccc|ccc|ccc}
\hline
\multirow{2}{*}{Merge Ratio} & \multicolumn{3}{c|}{Task} & \multicolumn{3}{c|}{Dare-Task} & \multicolumn{3}{c|}{Ties} & \multicolumn{3}{c}{Dare-Ties} \\
\cline{2-13}
 & Full-FT & $\boldsymbol{\mathcal{R}_{\textit{if}}^w}$ & ${\mathcal{R}_{\textit{if}}^l}$ & Full & ${\mathcal{R}_{\textit{if}}^w}$ & ${\mathcal{R}_{\textit{if}}^l}$ & Full & ${\mathcal{R}_{\textit{if}}^w}$ & ${\mathcal{R}_{\textit{if}}^l}$ & Full & $\boldsymbol{\mathcal{R}_{\textit{if}}^w}$ & ${\mathcal{R}_{\textit{if}}^l}$ \\
\hline
0.9:0.1 & 100.00\% & 100.00\% & 100.00\% & 100.00\% & 100.00\% & 100.00\% & 100.00\% & 0.00\% & 10.00\% & 100.00\% & 20.00\% & 20.00\% \\
0.8:0.2 & 100.00\% & 90.00\%  & 100.00\% & 100.00\% & 90.00\%  & 100.00\% & 100.00\% & 0.00\% & 10.00\% & 100.00\% & 0.00\%  & 20.00\% \\
0.7:0.3 & 100.00\% & 90.00\%  & 100.00\% & 100.00\% & 90.00\%  & 100.00\% & 100.00\% & 0.00\% & 0.00\%  & 100.00\% & 0.00\%  & 0.00\%  \\
0.6:0.4 & 100.00\% & 10.00\%  & 70.00\%  & 100.00\% & 0.00\%   & 70.00\%  & 100.00\% & 0.00\% & 0.00\%  & 100.00\% & 0.00\%  & 10.00\% \\
0.5:0.5 & 100.00\% & 0.00\%   & 0.00\%   & 100.00\% & 0.00\%   & 100.00\% & 100.00\% & 0.00\% & 0.00\%  & 100.00\% & 0.00\%  & 0.00\%  \\
\hline
\end{tabular}
\caption{Robustness under model merging on WizardMath using three fingerprinting strategies: Full Finetuning, Direct LoRA, and LoRA-FP Transferred.}
\label{tab:full_vs_lora_merge}
\end{table*}

\section{Cross-Model Family Generalization Validation}
\label{sec:app:cross}
To validate the generalizability of LoRA-FP across diverse LLM architectures, we conducted comprehensive experiments on multiple model families beyond our main LLaMA2-to-WizardMath case study.

\subsection{Transfer within LLaMA2 Family}
We extended our validation to a broader set of downstream models with architectural homology to LLaMA2-7B. Table~\ref{tab:fsr_llama2_family} shows the FSR for IF-based fingerprint transfer across four different LLaMA2-derived models, including Vicuna-7B-v1.5\footnote{\url{https://huggingface.co/lmsys/vicuna-7b-v1.5}}, WizardMath-7B\footnote{\url{https://huggingface.co/WizardLMTeam/WizardMath-7B-V1.0}}, Chinese-LLaMA2-7B\footnote{\url{https://huggingface.co/LinkSoul/Chinese-Llama-2-7b}} and LLaMA2-Finance-7B\footnote{\url{https://huggingface.co/cxllin/Llama2-7b-Finance}}. All transfers achieved 100\% success rate, confirming strong generalization within the model family.

\begin{table}[ht]
\centering
\scriptsize
\begin{tabular}{lc}
\hline
Downstream Model & FSR (IF from LLaMA2-7B) \\
\hline
Vicuna-7B-v1.5      & 100.00\% \\
WizardMath-7B       & 100.00\% \\
LLaMA2-Finance-7B   & 100.00\% \\
Chinese-LLaMA2-7B   & 100.00\% \\
\hline
\end{tabular}
\caption{\footnotesize FSR for LoRA-FP transfer within the LLaMA2 model family.}
\label{tab:fsr_llama2_family}
\end{table}

\subsection{Transfer Across Different Model Families}
To further validate cross-family transferability, we conducted experiments on several widely-used model families released recently, including Mistral-7B-v0.3\footnote{\url{https://huggingface.co/mistralai/Mistral-7B-v0.3}}, LLaMA3.1-8B\footnote{\url{https://huggingface.co/meta-llama/Llama-3.1-8B}}, and Qwen2.5-7B\footnote{\url{https://huggingface.co/Qwen/Qwen2.5-7B}} families. In each case, the fingerprint was injected into the original base model and then transferred to its downstream variant.

\begin{table}[ht]
\centering
\scriptsize
\begin{tabular}{lcc}
\hline
Original Model & Downstream Model & FSR \\
\hline
Mistral-7B-v0.3 & Mistral-7B-v0.3-Instruct & 100.00\% \\
LLaMA3.1-8B & LLaMA3.1-8B-Instruct & 100.00\% \\
Qwen2.5-7B & Qwen2.5-7B-Instruct & 100.00\% \\
Qwen2.5-7B & Qwen2.5-Math-7B & 100.00\% \\
\hline
\end{tabular}
\caption{\footnotesize LoRA-FP transfer effectiveness across different model families.}
\label{tab:fsr_cross_family}
\end{table}

As shown in Table~\ref{tab:fsr_cross_family}, LoRA-FP achieved 100\% FSR consistently, which demonstrates its reliable transfer of fingerprint signals both within a single model family and across distinct architectures.

\section{Detailed Analysis on Mistral Family}
\label{app:sec:mistral-family}
To demonstrate method generalizability beyond the LLaMA2 family, we conducted comprehensive evaluation on the Mistral-7B-v0.3 family. We embedded IF-based fingerprints on Mistral-7B-v0.3 and transferred them to Mistral-7B-v0.3-Instruct, comparing against direct fingerprint insertion.

\vspace{0.5em}
\paragraph{Harmlessness Evaluation} 
Table~\ref{tab:harmlessness_mistral} presents the harmlessness comparison between direct and transferred fingerprints across 19 benchmark datasets. The results show minimal variation between direct and transferred approaches, confirming that LoRA-FP introduces no additional harm to downstream utility compared to direct injection.

\begin{table}[ht]
\centering
\scriptsize
\begin{tabular}{lcc}
\hline
Dataset & Direct Fingerprint & Transferred from Mistral-7B-v0.3 \\
\hline
anli\_r1 & 0.457  & 0.455  \\
anli\_r2 & 0.431  & 0.440  \\
anli\_r3 & 0.4425 & 0.4433 \\
arc\_challenge & 0.6134 & 0.5861 \\
arc\_easy & 0.8417 & 0.7836 \\
openbookqa & 0.466  & 0.464  \\
winogrande & 0.7355 & 0.7434 \\
logiqa & 0.3471 & 0.3671 \\
sciq & 0.977  & 0.961  \\
boolq2 & 0.8412 & 0.8483 \\
cb & 0.4069 & 0.5882 \\
cola & 0.0723 & 0.1192 \\
rte & 0.6787 & 0.7328 \\
wic & 0.5517 & 0.5752 \\
wsc & 0.6538 & 0.5576 \\
copa & 0.9300 & 0.9300 \\
multirc & 0.4315 & 0.4011 \\
lambada\_standard & 0.6975 & 0.7100 \\
\hline
\end{tabular}
\caption{Harmlessness comparison between direct and transferred fingerprints on the Mistral family.}
\label{tab:harmlessness_mistral}
\end{table}

\vspace{0.5em}

\paragraph{Robustness Under Incremental Fine-tuning}

We evaluated robustness under incremental training using six different datasets. Table~\ref{tab:incremental_mistral} shows that while transferred fingerprints maintain reasonable robustness, there is some degradation compared to the LLaMA2 setting, though still within acceptable margins.

\begin{table}[ht]
\centering
\scriptsize
\begin{tabular}{lcc}
\hline
Dataset & Direct Fingerprint & Transferred from Mistral-7B-v0.3 \\
\hline
Alpaca-10k   & 100.00\% & 100.00\% \\
Alpaca-3k    & 100.00\% & 100.00\% \\
ShareGPT-6k  & 0.00\%   & 0.00\%   \\
ShareGPT-3k  & 30.00\%  & 20.00\%  \\
Dolly-10k    & 80.00\%  & 60.00\%  \\
Dolly-3k     & 100.00\% & 80.00\%  \\
\hline
\end{tabular}
\caption{Robustness under incremental fine-tuning on the Mistral family}
\label{tab:incremental_mistral}
\end{table}

\vspace{0.5em}

\paragraph{Robustness Under Model Merging}
We also evaluated robustness under four different model merging strategies, using OpenLLM-Ro/RoMistral-7B-Instruct as the second expert model. Table~\ref{tab:merge_mistral} shows the results across different merging ratios. While fingerprints transferred to Mistral still performed robustly, there was slight degradation compared to the LLaMA2 setting (10-20\% drop), though remaining within acceptable margins.


\begin{table*}[t] 
\centering
\scriptsize
\setlength{\tabcolsep}{6pt}
\renewcommand{\arraystretch}{1.1}

\resizebox{\textwidth}{!}{%
\begin{tabular}{c|cc|cc|cc|cc}
\hline
\multirow{2}{*}{Rate} & \multicolumn{2}{c|}{Task} & \multicolumn{2}{c|}{Dare-Task} & \multicolumn{2}{c|}{Ties} & \multicolumn{2}{c}{Dare-Ties} \\
\cline{2-9}
 & Direct & Transferred & Direct & Transferred & Direct & Transferred & Direct & Transferred \\
\hline
0.9:0.1 & 100.00\% & 100.00\% & 100.00\% & 100.00\% & 100.00\% & 100.00\% & 100.00\% & 100.00\% \\
0.8:0.2 & 100.00\% & 100.00\% & 100.00\% & 100.00\% & 100.00\% & 100.00\% & 100.00\% & 100.00\% \\
0.7:0.3 & 100.00\% &  90.00\% & 100.00\% &  90.00\% & 100.00\% & 100.00\% & 100.00\% & 100.00\% \\
0.6:0.4 &  80.00\% &  60.00\% & 100.00\% &  60.00\% & 100.00\% &  70.00\% & 100.00\% &  80.00\% \\
0.5:0.5 &  50.00\% &  30.00\% &  50.00\% &  30.00\% &  70.00\% &  50.00\% &  80.00\% &  60.00\% \\
\hline
\end{tabular}%
}

\caption{Robustness under model merging on the Mistral family: comparison between direct fingerprint insertion and transferred fingerprints across different merging strategies.}
\label{tab:merge_mistral}
\end{table*}

\section{Comparison with Full-Parameter Fine-tuning}
\label{app:sec:full}
To further evaluate the effectiveness of LoRA-FP, we conduct a comprehensive comparison against direct fingerprint injection using full-parameter fine-tuning (Full-FT), using the IF fingerprinting method as a representative example. The comparison focuses on three key dimensions: harmlessness, robustness, and computational efficiency.

\paragraph{Harmlessness}  
As previously reported in Table~\ref{tab:ones_table}, LoRA-FP generally leads to better preservation of downstream task performance compared to Full-FT. This highlights the advantage of LoRA-FP in maintaining the model’s utility after fingerprint injection.

\paragraph{Robustness}  
We evaluate robustness from two perspectives: model merging~(Table~\ref{tab:full_vs_lora_merge}) and incremental fine-tuning~(Table~\ref{tab:full_ft_downstream}). While full fine-tuning achieves the highest fingerprint retention across different merging strategies, it comes at the cost of greater degradation in harmlessness. In contrast, LoRA-FP transferred fingerprints exhibit significantly stronger robustness than direct LoRA injection and retain high fingerprint success rates, often approaching the level achieved by full fine-tuning.

\begin{table}[H]
\centering
\scriptsize
\resizebox{\linewidth}{!}{
\begin{tabular}{l|c|c|c|c}
\hline
\textbf{Method} & \textbf{Downstream Dataset} & \textbf{Full-FT on WizardMath} & $\boldsymbol{\mathcal{R}_{\textit{if}}^w}$ & $\boldsymbol{\mathcal{R}_{\textit{if}}^l}$ \\
\hline
\multirow{6}{*}{IF} 
 & Alpaca-10k     & 100.00\% & 0.00\%  & 50.00\% \\
 & Alpaca-3k      & 100.00\% & 0.00\%  & 70.00\% \\
 & ShareGPT-6k    & 0.00\%   & 0.00\%  & 0.00\% \\
 & ShareGPT-3k    & 0.00\%   & 0.00\%  & 0.00\% \\
 & Dolly-10k      & 100.00\% & 0.00\%  & 50.00\% \\
 & Dolly-3k       & 100.00\% & 0.00\%  & 90.00\% \\
\hline
\end{tabular}
}
\caption{Comparison between direct full-parameter fingerprint injection and LoRA-FP transfer on WizardMath using the IF method.}
\label{tab:full_ft_downstream}
\end{table}

\paragraph{Computational Cost}  
In terms of training efficiency, LoRA-FP offers substantial advantages. Since it only updates small adapter modules rather than full model parameters, it avoids the overhead of complete model storage and retraining. Although explicit measurements are not included in the main text, empirical observations indicate that LoRA-FP training can be completed within minutes using less than 24GB of memory. In comparison, full-parameter fine-tuning typically requires around one hour and 64GB of memory (with half-precision), making LoRA-FP a more scalable and practical option for deployment in resource-constrained environments.

Overall, while full fine-tuning offers slightly better robustness under extreme conditions, LoRA-FP provides significant advantages in terms of training efficiency and harmlessness, making it a more practical and scalable fingerprint solution for real-world deployment.
\end{document}